\documentclass[article,12pt,nofootinbib]{revtex4}
\usepackage[paper=letterpaper,margin=1.5in]{geometry}
\usepackage{epsfig,color}
\usepackage{graphicx}
\usepackage{amssymb,amsmath}
\usepackage{time}
\usepackage[varg]{txfonts}
\usepackage{hyperref}
\hypersetup{
  colorlinks,
  citecolor=green,
  linkcolor=blue}
\newcommand{\be}{\begin{equation}}
\newcommand{\ee}{\end{equation}}
\newcommand{\bc}{\begin{center}}
\newcommand{\ec}{\end{center}}

\begin{document}

\bibliographystyle{revtex}


\vspace{.4cm}

\title
{Accurate  measurement of  uncorrelated energy spread in  electron beam
}
\author{Sergey Tomin, Igor  Zagorodnov, \\ Winfried Decking, Nina Golubeva and Matthias Scholz}
\affiliation{Deutsches Elektronen-Synchrotron, Notkestrasse 85,
22603 Hamburg, Germany\vspace{.4cm}}
\date{\today}

\vspace{.4cm} 
\begin{abstract} 
We present measurements of slice energy spread at the injector section of the European
X-Ray Free Electron Laser for an electron bunch with charge of 250 pC.  Two methods considered  in the paper are based on measurements at the dispersive section after a transverse deflecting structure (TDS). The first approach uses measurements at different beam energies. We show that with a proper scaling of the TDS voltage with the beam energy the rms error of the measurement is less than 0.3 keV for the energy spread of 6 keV.  In the second approach we demonstrate that keeping the beam energy constant but adjusting only the optics we are able to simplify the measurement complexity and to reduce the rms error below 0.1 keV. The accuracy of the measurement is confirmed by numerical modelling including beam transport effects and collective beam dynamics of the electron beam. The slice energy spread measured at the European XFEL for the beam charge of 250 pC is  nearly 3  times lower  as the one reported recently at SwissFEL for the same cathode material and the beam charge of 200 pC.

\end{abstract}

\maketitle

\section{Introduction}\label{sec:1}

 The small emittance and the low energy spread of the electron beam required at X-Ray Free Electron Lasers (XFELs) can cause the microbunching  instability~\cite{MB1, MB2} and destroy the lasing.  On the other hand, a large initial  energy spread  will hinder  a proper compression of the bunch and will lead to intolerable  energy spread after compression. Hence a reliable high resolution method of the measurements of the uncorrelated (or slice) energy spread is crucial for a proper operation of the modern facilities.
 
In order to measure the slice energy spread a standard approach with a transverse deflector and the dispersive section is used. However, it shows only a low resolution (of several keV) due to  impact of OTR screen resolution, the betatron beam size and the deflector strength on the measurement. The energy spread induced by deflector can be excluded with a set of measurements with different  deflector amplitudes. Such experiments have been done at PITZ~\cite{MK20}. 

Recently in~\cite{Prat20}  it was suggested to carry out the measurements at different electron beam energies. For the setup used at SwissFEL - deflector at constant energy and an acceleration section  after it -  the authors have written  a polynomial equation of the second order and analyzed the accuracy of the coefficient reconstruction relative to statistical and systematic errors.  On the basis of this analysis they concluded that the accuracy of the measurements for their setup could be better than 1 keV.  In order to exclude the impact of the deflector the authors in~\cite{Prat20}   suggested to carry out additional sets of measurements, but it was not done and an analytical estimation was used instead. 

The energy scan method seems to be simple. However at the layout of the European XFEL this method requires modifications and even with the changes described in this paper it has not shown the expected performance.

\begin{figure}[htbp]
	\centering
	\includegraphics*[height=35mm]{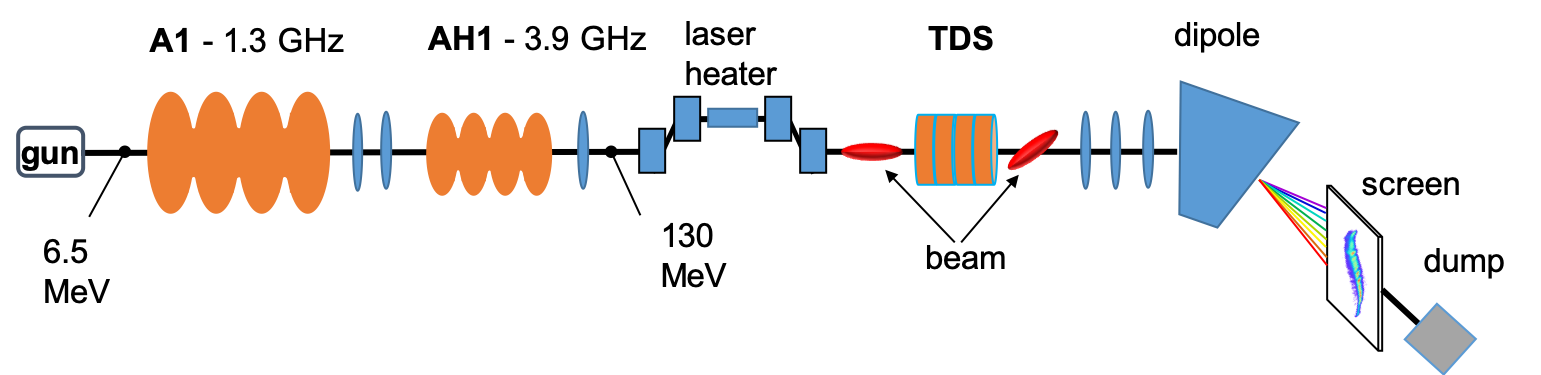}
	\caption{The setup of the experiment at the injector section of the European XFEL.}\label{Layout}
\end{figure}

At this paper we present the measurement at the European XFEL with layout shown in Fig.~\ref{Layout}. The situation is different as compared to SwissFEL case. The deflector is placed after the acceleration section, and the beam energy in it changes during the experiment. This case was mentioned in~\cite{Prat20} too and it was suggested to use a polynomial equation of the third order with additional sets of measurements in order to exclude the impact of the deflector.

If we use only energy scan then in our setup  the accuracy of the reconstruction of the polynomial of the third order is low and very sensitive to errors.  In order to overcome it we suggest a proper scaling of the deflector voltage with the beam energy. It allows (1) to reduce the order of the polynomial equation; (2) to avoid a need in additional sets of measurement with different deflector amplitudes. And the most important  advantage of the suggested method is (3) a considerable higher accuracy of the reconstruction of the coefficients and as a consequence a higher accuracy of the energy spread measurements. 

The experiments described in our paper are done at low energetic part of the facility.  In order to analyze the accuracy of the measurement we have done numerical modeling including beam transport effects and collective beam dynamics of the electron beam. We have found that the radio-frequency (RF) focusing impacts the beam considerably and a matching of the beam to the optics before the TDS is necessary for each beam energy. Additionally we see in the modeling and in the experiment that due to collective effects the emittance and the energy spread are not constant during the energy scan.

Applying the method based on the energy scan in  practice we have found that  (for our setup and a small interval of energy change available) the method is very time-intensive due to beam matching and the optics scaling. In the measurements we have failed to obtain reliable data which allow to carry out an accurate reconstruction. Hence we developed another method and have used the data from energy scan experiments only to show the consistence with the results obtained by the second method.

In the second approach we keep the beam energy constant and avoid time-consuming matching of the beam before TDS. By adjusting the strength of quadrupoles after TDS we are able to carry out independent scans in dispersion,  in TDS strength and in beta-function on OTR screen. The method is fast and allows to obtain accurate measurements of slice energy spread with resolution better than 0.1 keV. 

The paper is organized as follows. The methods of the measurement and their analysis are described in Section~\ref{sec2}. The beam dynamics modeling of the approaches with collective effects is considered in Section~\ref{sec3}.  Then, in Section~\ref{sec4} the results of the measurements at the European XFEL injector and their analysis are presented. 

\section{Methods of the measurement and their analysis}\label{sec2}
For the setup of Fig.~\ref{Layout} the measured beam size $\sigma_M$ on the screen can be written as
\begin{align}\label{Eq_model}
	\sigma_M^2= \sigma_R^2+\frac{E_0}{E}\sigma_B^2+\frac{D^2 }{E^2}\sigma_E^2+\frac{(D e k V)^2 E_0 }{E^3}\sigma_I^2,\\
	\sigma_B^2=\frac{\beta_x \epsilon_n}{\gamma_0}, \quad \sigma_I^2=\frac{\epsilon_n (\beta_y^{0}+0.25 L^2\gamma_y^{0}-L\alpha_y^{0})}{\gamma_0},\nonumber
\end{align}	
where $E$ is the beam energy, $\sigma_R$ is the screen resolution, $\beta_x$ is the optical function at the position of the screen,  $\epsilon_n$  is the normalized beam emittance, $\gamma_0$ is the relative beam energy, $D$  is the horizontal dispersion at the screen position; $\beta_y^{0}$,  $\gamma_y^{0}$ and $\alpha_y^{0}$ are the twiss parameters at the beginning of TDS; $k$, $V$, and $L$ are the wave number, voltage and length of the TDS and $e$ is the electron charge.  

\begin{table}[htbp]
	\centering
	\caption{Simulation parameters.}
	\label{Table_SP}
	\begin{tabular}{lcccll}
		\hline\hline
		{\bf parameter}& 	{\bf Units}& 	{\bf Value}\\
		\hline
		OTR resolution, $\sigma_R$  &$\mu$m&28\\
		Normalized emittance, $\epsilon_n$  &$\mu$m&0.4\\
		Reference optical  $\beta$-function at OTR , $\beta_x^0$  &m&0.6\\
		Reference dispersion,  $D_0$  &m&1.2\\
		Optical $\beta$-function at TDS,  $\beta_y^0$  &m&4.3\\
	    Optical $\alpha$-function at TDS,  $\alpha_y^0$  & &1.9\\
		Wave number of TDS,  $k$  & 1/m&58.7\\
		Length of TDS,  $L$  & m&0.7\\
		Reference voltage of TDS,  $V_0$  & MV&0.61\\
		Reference energy,  $E_0$  & MeV&130\\
		\hline\hline
	\end{tabular}
\end{table}

In this section we propose and validate two methods to measure the slice energy spread.  The parameters used in the simulations are  listed in Table~\ref{Table_SP}. These parameters are close to estimations obtained in the experiment. The resolution of OTR screen is 28 $\mu$m and it  agrees with other publications: 10-20 $\mu$m in ~\cite{OTR}, 30 $\mu$m in ~\cite{Prat20}.

\subsection{Method based on energy scan}\label{sec2.1}

If the second and the third terms in the right hand side of  Eq.(\ref{Eq_model}) are not known
then the energy spread can be estimated as
\begin{align}\label{}
	\sigma_E \approx\frac{E}{D}\sigma_M.
\end{align}	
The error of this estimation is defined by the resolution
\begin{align}\label{}
R_{\sigma_E}=\frac{E}{D}\sqrt{\sigma_R^2+\frac{E_0}{E}\sigma_B^2 +\frac{(D e k V)^2 E_0 }{E^3}\sigma_I^2}.
\end{align}

In order to increase the resolution it was suggested in~\cite{Prat20} to "perform beam size measurements for different energies and deflector voltages and to fit the data"  with Eq.(\ref{Eq_model}). Note that Eq.(\ref{Eq_model}) uses a more accurate approximation of the last term compared to ~\cite{Prat20} where the authors used  $\sigma_I^2=\epsilon_n \beta_y^0/\gamma_0$.

If we keep the voltage of the deflector constant and change only the beam energies than we can fit the measurements to Eq.(\ref{Eq_model}) in hope to reconstruct all coefficients of this polynomial.  We simulated with Eq.(\ref{Eq_model}) a measurement of the beam size  $\sigma_M$  with constant TDS voltage $V_0$ and the beam energy changing between 90 and 190 MeV with step of 10 MeV. At each beam energy we simulate 30 measurements  of the beam size  $\sigma_M$ with random error of $2\%$. We consider the slice energy spread between 0.5 and 7 keV. In the fit we used the  simplex search method of Lagarias et al~\cite{Fit}.

From numerical experiment we have found that the rms error of the reconstruction of energy spread is larger than 2 keV. Under the rms error of reconstruction in the paper we mean the value  defined as
\begin{align}\label{}
	\Delta_{\sigma_E}=\sqrt{\frac{1}{N}\sum_{i=1}^N (\sigma_E-\sigma_E^0)^2},
\end{align}
where N is the number of shots (reconstructions), $\sigma_E$ is the energy spread obtained from the reconstruction (of the polynomial coefficient from the simulated measurements) and  $\sigma_E^0$ is true energy spread used in the simulation of the reconstruction procedure.  In order to estimate this error we used 100 shots at each energy spread point.

In order to reduce the error we can do an additional scan with different deflector voltages to estimate the last term in Eq.(\ref{Eq_model}). With this estimation we reduce  the error of the reconstruction. However, we will not  analyze this approach here and suggest below another technique to reduce the order of the polynomial and to increase the accuracy of the reconstruction of the polynomial coefficients. 

It can be achieved if we will keep constant not the voltage $V$ but the streaking parameter of the deflector:
\begin{align}\label{}
S_0=\sqrt{\beta_y \beta_y^{0}} \sin(\Delta\mu_y) K_0,\quad K_0=\frac{e V_0 k}{E_0},
\end{align}	
where $\Delta\mu_y$ is the phase advance between the middle of TDS and the OTR screen, $\beta_y^0$ is the optical function at the   TDS,   $\beta_y$ is the optical function at the position of  OTR  and the voltage $V_0$ is a fixed value which produces the desired streak $S_0$ at the fixed beam energy $E_0$.		

In the following we adjust the voltage of TDS  proportionally to the beam energy:
\begin{align}\label{Eq_V}
	V(E)=\frac{V_0}{E_0} E.
\end{align}

\begin{figure}[htbp]
	\centering
	\includegraphics*[height=110mm]{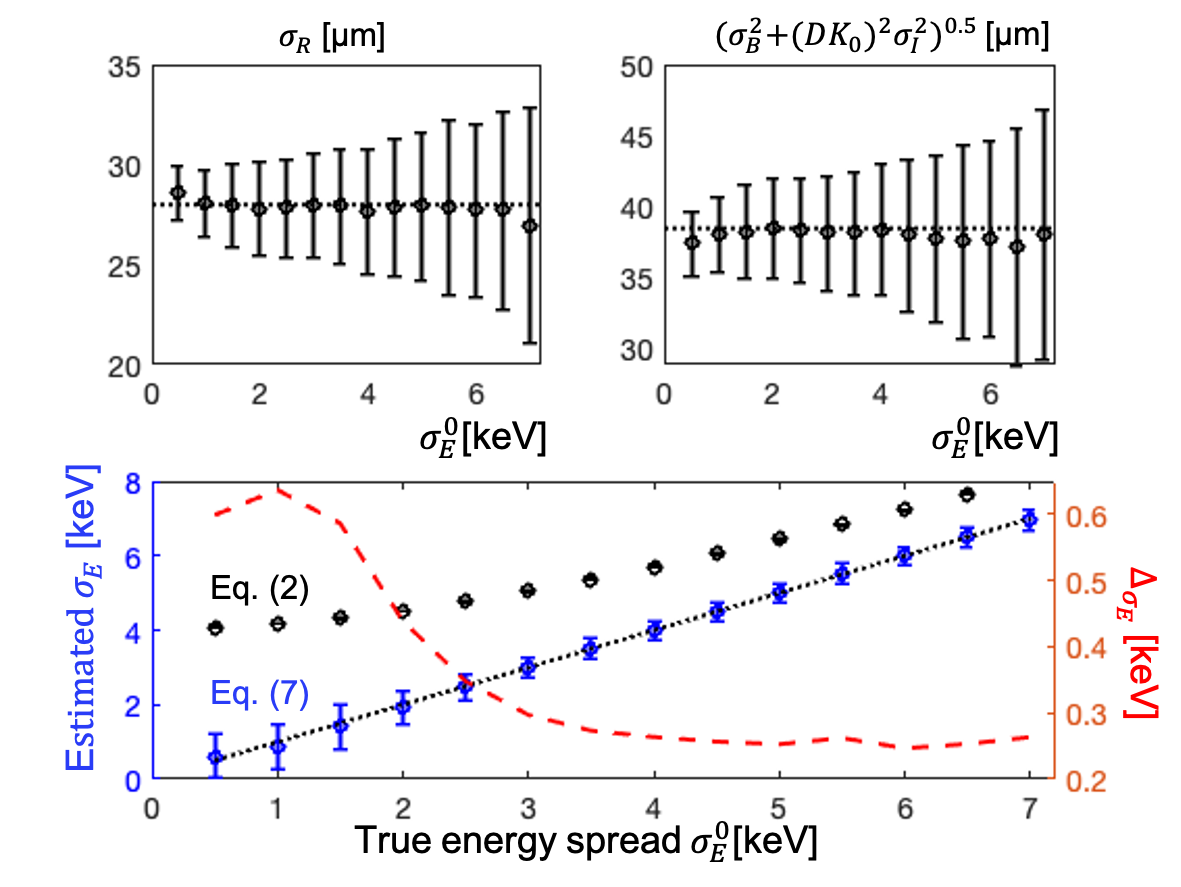}
	\caption{Performance of the method based on energy scan. Circles shows reconstructed mean values. Error bars give the standard deviation error. The red dashed curve presents the rms error of the reconstruction. Dotted black curves show true values used for the simulation.}\label{Fig_Ex2}
\end{figure}

If we put Eq.(\ref{Eq_V})  in Eq.(\ref{Eq_model}) then we reduce the order of the polynomial from the third to the second one: 
\begin{align}\label{eq_main}
	\sigma_M^2= \sigma_R^2+\frac{E_0}{E}\sigma_{BI}^2+\frac{D^2}{E^2}\sigma_E^2,\quad
	\sigma_{BI}^2=\sigma_B^2+(D K_0\sigma_I)^2.
\end{align}	

We simulated with Eq.(\ref{eq_main}) a measurement of the beam size  $\sigma_M$  with the beam energy changing between 90 and 190 MeV with step of 10 MeV.  We used the same errors and the reconstruction algorithm as in the previous example. The results of the reconstruction are shown in Fig.~\ref{Fig_Ex2} and the error of the reconstruction of energy spread is nearly 0.3  keV at the energy spread of 6 keV. 

In experiment and in  the beam dynamics simulations we have not been able to show this accuracy. We show with beam dynamics simulations that in order to use this method we have to match the beam to the optics at each beam energy. It requires considerable efforts and we failed to make it with high accuracy at the experiment. Additionally, in the modelling and the experiment we have seen that the slice emittance is not constant.

\subsection{Method based on dispersion scan}\label{sec2.2}

In this section we present another method which use constant beam energy $E_0$ and avoids above described difficulties. The method shows much better resolution  theoretically and it  is easy to use experimentally.

We have developed a special optics described in the next section. Using only few quadrupoles between TDS and the OTR screen we are able to  change the dispersion $D$ at the OTR position keeping $\beta_x$ -function  constant with only moderate changes in $\beta_y$-function  and in the streaking $S$. 

We start with changing of TDS voltage $V$ and fit the measured slice size $\sigma_M$ to the quadratic polynomial:
\begin{align}\label{eq_sV}
	\sigma_M^2= A_V+B_V V^2.
\end{align}	
During the scan we keep the dispersion at constant value $D_0$.

\begin{figure}[htbp]
	\centering
	\includegraphics*[height=110mm]{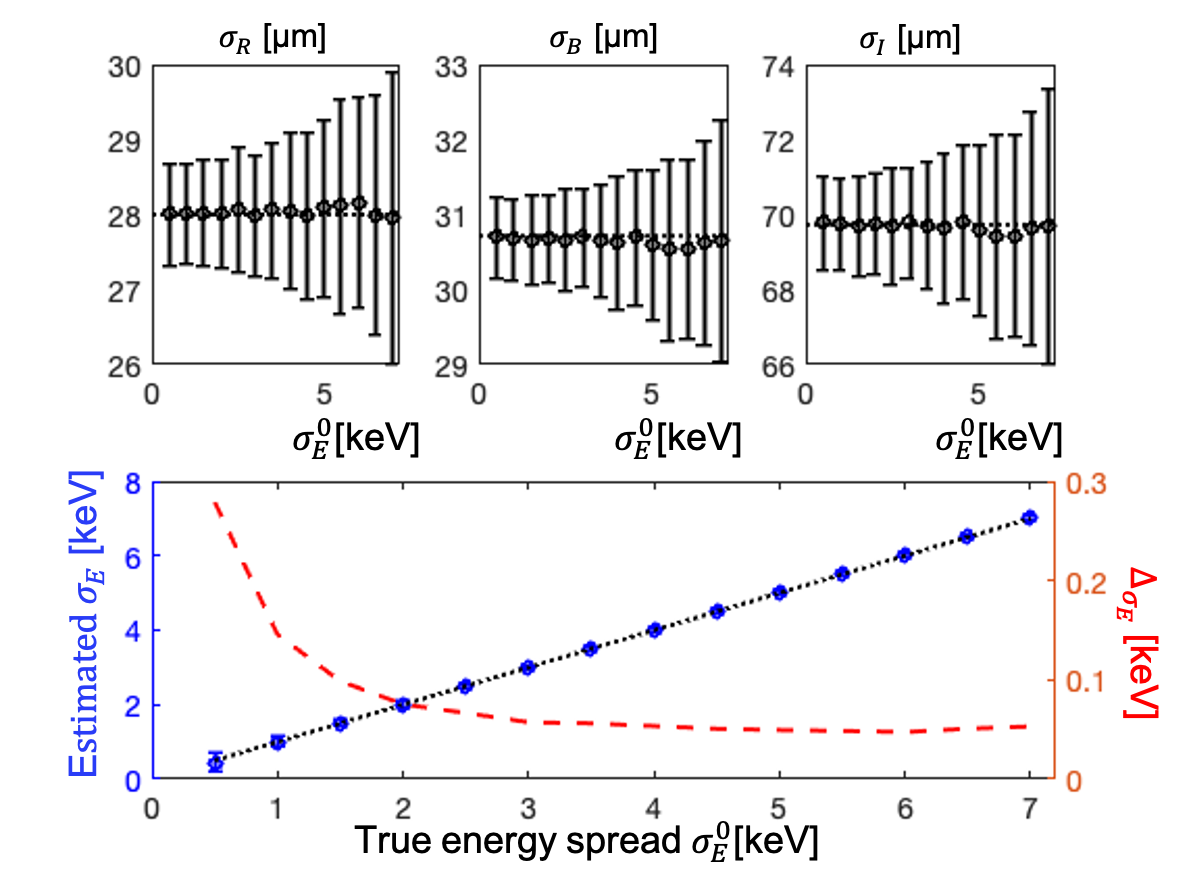}
	\caption{Performance of the method based on dispersion scan. Circles shows reconstructed mean values. Error bars give the standart deviation error. The red dashed curve presents the rms error of the reconstruction. Dotted black curves show true values used for the simulation.} \label{Fig_Ex3}
\end{figure}

At the second step we keep constant the TDS voltage  at $V_0$ and change the dispersion $D$. We  fit the measured slice size $\sigma_M$ to the quadratic polynomial:
\begin{align}\label{eq_sD}
	\sigma_M^2= A_D+B_D D^2.
\end{align}	

After these two fits we are able to find out all terms of Eq.(\ref{Eq_model}):
\begin{align}\label{eq_sall}
	\sigma_E= \frac{E_0}{D_0}\sqrt{A_D-A_V}, \quad 
	\sigma_I= \frac{E_0}{D_0 e k}\sqrt{B_V},\\
	\sigma_B= \sqrt{B_{\beta} \beta_x^0},\quad
	\sigma_R= \sqrt{A_D-\sigma_B^2},\nonumber
\end{align}
where 
\begin{align}\label{eq_sB0}
	B_{\beta}=\sigma_I^2(\beta_y^0+0.25 L^2 \gamma_y^0-L \alpha_y^0)^{-1}.
\end{align}

Eq.~(\ref{eq_sB0}) calculates the coefficient $B_{\beta}$ from the results of the TDS voltage scan, Eq.~(\ref{eq_sV}) . Otherwise,  if we had measured the slice emittance $\epsilon_n$ independently, then we can use more accurate estimation of  $B_{\beta}$ through the relation $B_{\beta}=\epsilon_n/\gamma_0$ . For example, we can estimate $B_{\beta}$ (or emittance $\epsilon_n$) changing only $\beta_x$ function at the OTR screen position and keeping the dispersion $D$ constant and fitting the measured slice size $\sigma_M$ to the linear polynomial:
\begin{align}\label{eq_sB}
	\sigma_M^2= A_{\beta}+B_ {\beta} \beta_x.
\end{align}	

We simulated with Eq.~(\ref{Eq_model}) the measurement of the beam size  $\sigma_M$  for   two scans as given by Eqs.(\ref{eq_sV})-(\ref{eq_sB0}). For the dispersion scan we used  the values of 0.6, 0.8, 1.0 and 1.2 meters. For the TDS voltage scan we used values 0.38, 0.47, 0.56, 0.65 and 0.75 MV.  We used the same errors and the reconstruction algorithm as in the previous examples. The results of the reconstruction are shown in Fig.~\ref{Fig_Ex3} and the error of the reconstruction of energy spread is smaller than 0.1  keV at the energy spread of 6 keV. 

\subsection{Impact of systematic and random instrumental errors}\label{sec2.3}

Finally, let us consider instrumental errors in the setup of TDS voltage $V$ and the dispersion $D$ during the scans used in the dispersion scan method. 

\begin{figure}[htbp]
	\centering
	\includegraphics*[height=80mm]{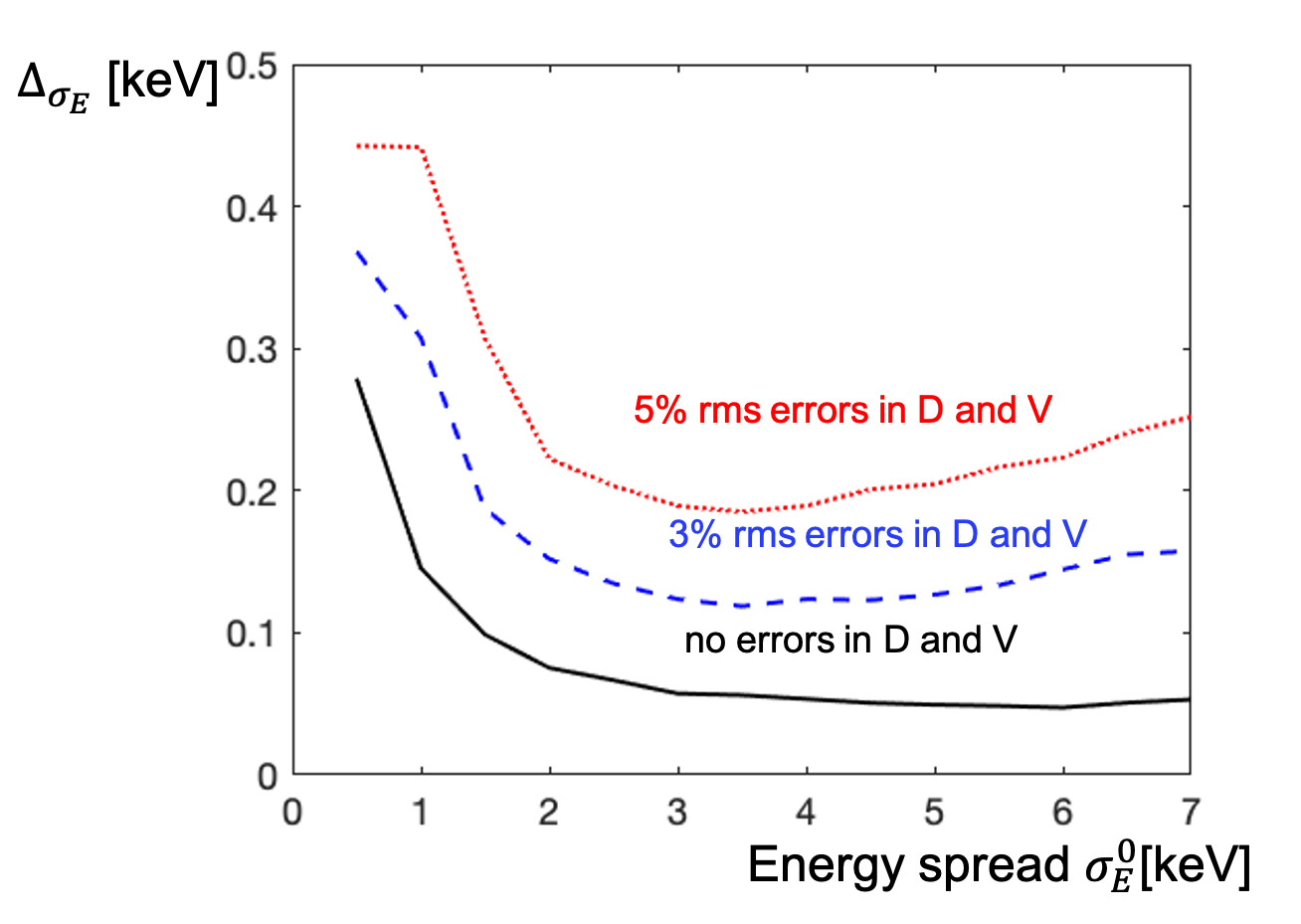}
	\caption{Impact of instrumental errors in setup of voltage and dispersion on the reconstruction error from dispersion scan method.} \label{Fig_Ex4}
\end{figure}

If the errors are systematic with the same sign then the reconstruction of energy spread only weakly affected by them. Indeed, we calculate energy spread by Eq.~(\ref{eq_sall})  and use only the constant terms $A_D$ and $A_V$. If we suggest that during the TDS voltage scan we set the voltage with the same negative  error, for example it is 10 \%, then it has only impact on coefficient   $B_V$ which in this case will be increased by factor $0.9^{-2}$, but the constant term $A_V$ is not changed. The same is true for the impact of the systematic error in the dispersion $D$ during the dispersion scan. 

Hence we analyze only  random errors in the setup of voltage or dispersion. The results of the analysis are shown in Fig.~\ref{Fig_Ex4}. In the suggestion of rms error of 5\% the rms reconstruction error remains below 0.3 keV for the energy spread of 6 keV.

\section{Modelling of the experiment with collective effects and the beam transport}\label{sec3}

The electron beam dynamics at the European XFEL accelerator has been recently discussed in  ~\cite{Zag19} and ~\cite{Zag20}. In the last work, an experimental validation of the collective effects modeling at the European XFEL injector was presented.  Here we use the same approach from ~\cite{Zag20}  to simulate the beam dynamics,  namely (1) the dynamics of the electron beam in the gun was simulated using  ASTRA code ~\cite{ASTRA},  (2) the beam tracking starting from 3.2 m from the gun cathode, was performed using Ocelot code ~\cite{OCELOT} with the space charge  and the wakefield effects included, (3) the coherent synchrotron radiation  was omitted as negligible for this section. 

\begin{figure}[htbp]
	\centering
	\includegraphics*[width=.9\textwidth]{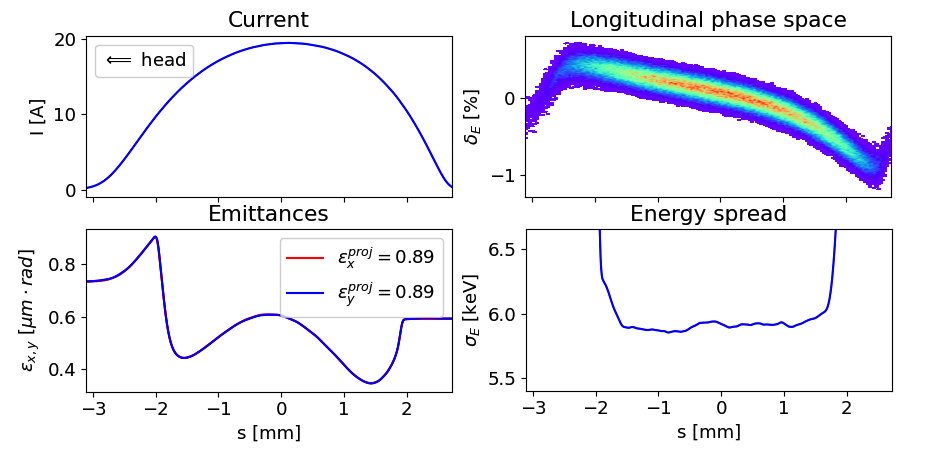}
	\caption{ Electron beam distribution after the gun used in the modeling.} \label{Fig_gun_distrib}
\end{figure}

In the measurements described in Section~\ref{sec4} we have found that the uncorrelated energy spread is equal to approximately 5.9 keV. As it is discussed in Section~\ref{sec5} one of the possible reasons of such large energy spread could be intrabeam scattering (IBS). The codes ASTRA and Ocelot do not model  IBS. In the simulation of RF gun with ASTRA for charge of 250 pC we obtain the energy spread of 0.6 keV. Hence we apply random generator at distance of 3.2 m from the cathode to increase the energy spread artificially to 5.9 keV.  The properties of the electron bunch after this procedure at position $z=3.2$ m are shown in Fig.~\ref{Fig_gun_distrib}. Let us note here that the projected emittance at this distance from the cathode is relatively large. The emittance will reduce in the booster considerably according with the emittance compensation process~\cite{Carlsten}.

\subsection{Magnetic lattice and its properties}\label{sec3.1}

\begin{figure}[htbp]
	\centering
	\includegraphics*[width=0.8\textwidth]{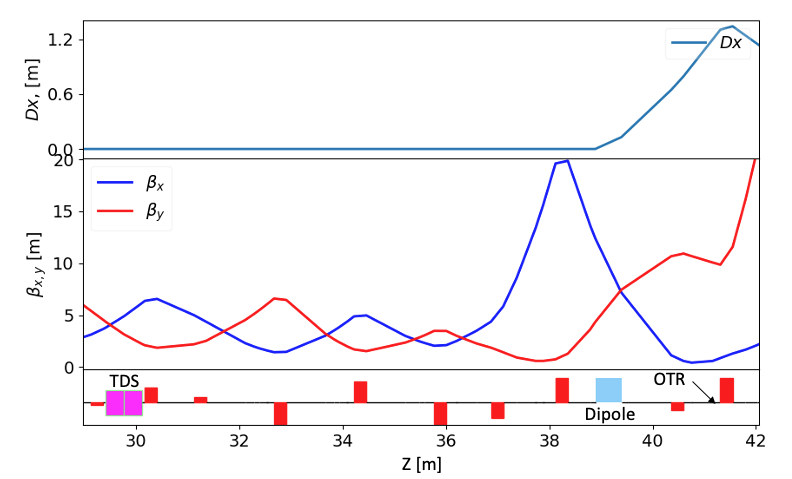}
	\caption{Special optics for the slice energy spread measurement. Optics is shown from the matching point.  Note, $\beta$-functions are calculated in linear approximation without any collective effects.} \label{Fig_Ex5}
\end{figure}

A special optics (shown in  Fig.~\ref{Fig_Ex5} and Table~\ref{Table_SP})  was developed to fulfil requirements of the experiments described above. The important properties of the optics are: (1) high dispersion at the screen position to maximize energy spread contribution to the beam size on the screen; (2) the high value of $R_{34}$ element of the transport matrix  between the TDS and the screen to minimize voltage of the TDS which in turn minimizes the induced energy spread; (3) low sensitivity of $\beta_x$ in the screen position to the beam mismatch at the matching point before TDS.

\begin{figure}[htbp]
	\centering
	\includegraphics*[width=0.8\textwidth]{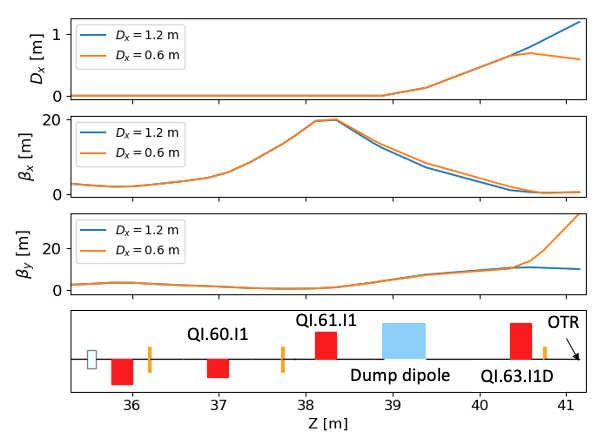}
	\caption{Changes of the twiss parameters in the dump section during dispersion scan.} \label{Fig_twiss_disp_scan}
\end{figure}

Another feature of this optics is the ability to vary the horizontal dispersion over a wide range without change in the horizontal $\beta$-function on the OTR screen. The main contribution for the dispersion change comes from the quadrupole QI.63.I1D while minor changes in the $\beta_x$-function are compensated by quadrupoles QI.60.I1 and QI.61.I1, Fig~\ref{Fig_twiss_disp_scan}. The described feature is used to performed the dispersion scan. 

\subsection{Numerical modelling of the experiment based on the energy scan}\label{sec3.2}

Following the measurement procedure described in the Section \ref {sec2.1},  we performed beam dynamics simulations changing the voltage  of module A1 (see Fig.~\ref{Layout}) from  80 to 180 MV with step 10 MV.  The beam comes from the gun with initial energy of 6.5 MeV. 

\begin{figure}[htbp]
	\centering
	\includegraphics*[width=1.0\textwidth]{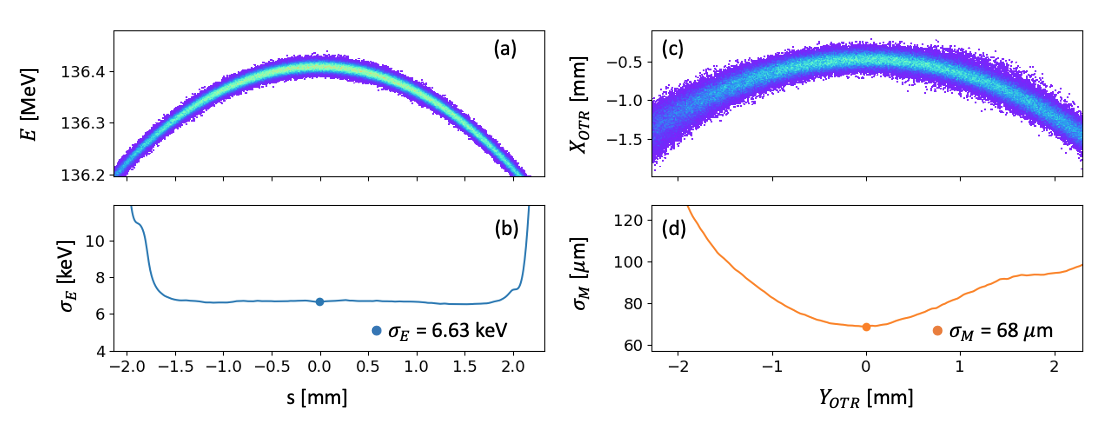}
	\caption{Some details of longitudinal beam dynamics for the beam energy 136 MeV.   \textbf{a)} The LPS beam distribution in front of the dump magnet, \textbf{b)} slice energy spread of the beam in front of dump magnet, \textbf{c)} the beam image on the OTR screen and  \textbf{d)} horizontal slice beam size on the OTR screen without effect of the screen resolution.} \label{Fig_LPS}
\end{figure}

The transport matrix from TDS to the screen in the dump section has non-zero $R_{51}$, $R_{52}$ elements (contribution form the dump magnet). These elements create couplings between the horizontal and longitudinal planes, which leads to the expansion of the beam in the longitudinal direction in case of non zero horizontal emittance.  In the presence of the correlated energy spread it will cause of the slice widening on the OTR screen. This effect in simulations can be see in  Fig.~\ref{Fig_LPS}. On the left are two plots of the LPS beam distribution (a) and the slice energy spread (b) of the beam in front of the dump magnet, and on the right two plots are the image (c) and the horizontal slice size (d) on the OTR screen of the same beam that was tracked through the dump section. To avoid the influence of this effect on the measurement, the beam slice of interest must have zero energy chirp. For completeness, it is worth noting that the element $R_ {56}$ is small and its effect can be neglected. 

\begin{figure}[htbp]
	\centering
	\includegraphics*[width=.8\textwidth]{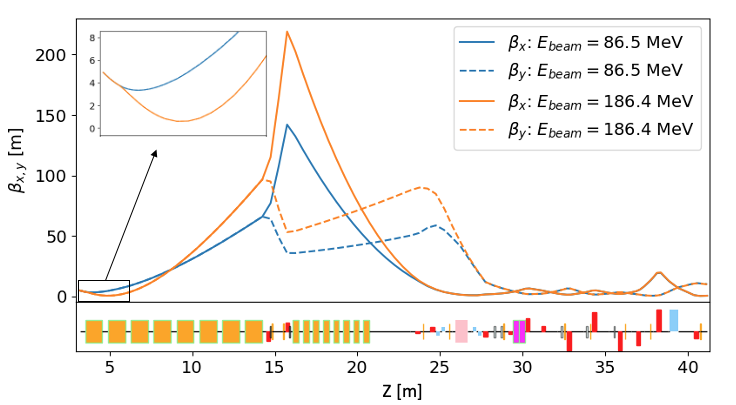}
	\caption{ $\beta$-functions for lowest and highest energy with taking into account the SC effect and RF focusing.} \label{Fig_match}
\end{figure}

Taking into account the effect described above, the third harmonic cavity AH1 was turned off.  measurement of the slice energy spread was carried out at the extremum of the mean slice energy. The horizontal twiss parameters of the slice have been matched to the magnetic lattice before TDS. Since the RF-focusing effect is strong, the beam should be matched for all beam energies. At the Fig.~\ref{Fig_match} are shown $\beta$-functions of the central slice for highest and lowest energies. Twiss parameters were calculated from the beam transported in Ocelot with collective effects included.  

In the simulations and in the experiment we have seen increase of the slice emittance by 30\% at the highest voltage of RF module A1. It is due to very strong RF focusing and very small $\beta$-functions in module A1 which, in turn, enhance the SC effect. Additionally we think that IBS would change  the energy spread during the energy scan as well. But this effect was omitted in the simulation.

The true values used in the simulation are listed in the first row of Table~\ref{T2}.

\begin{figure}[htbp]
	\centering
	\includegraphics*[width=0.95\textwidth]{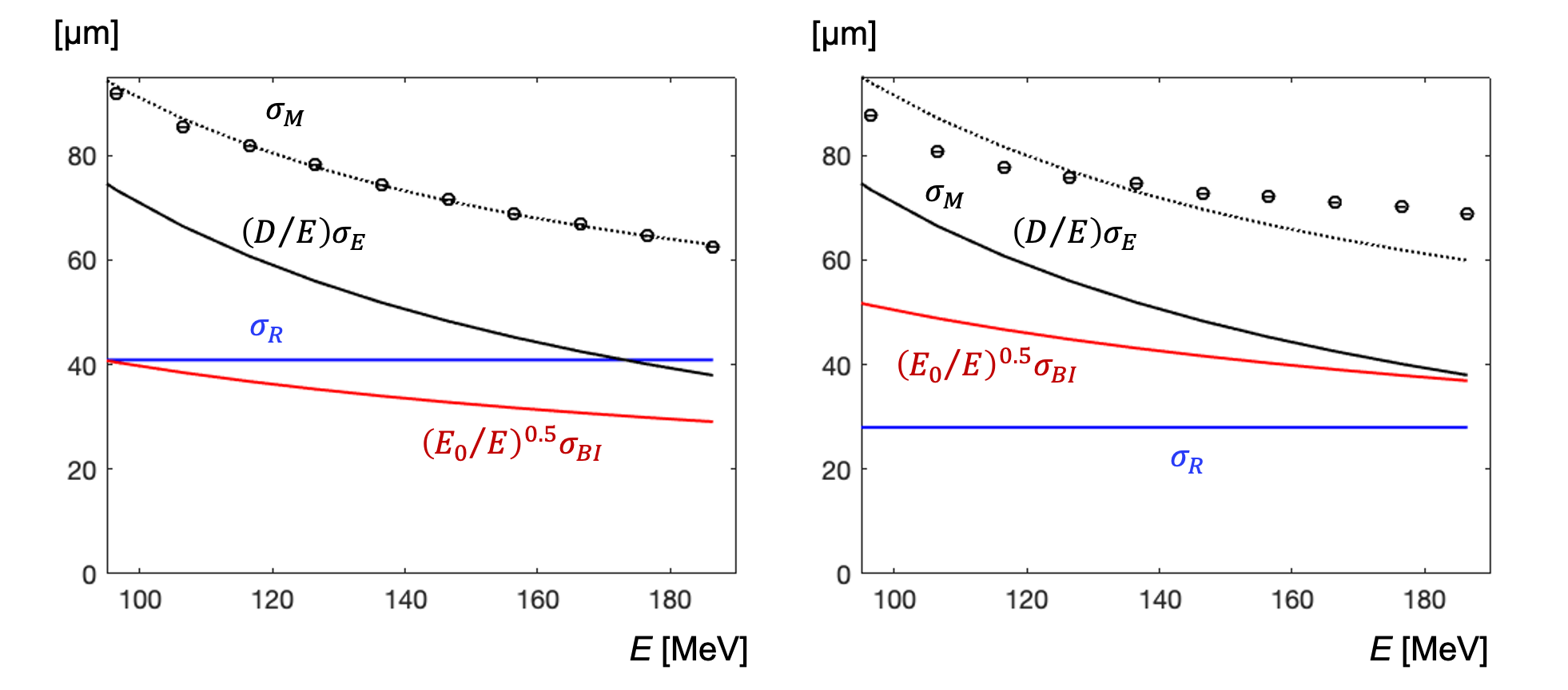}
	\caption{The left plot shows the results for the beam matched at each energy. The right plot shows the results for the beam matched only at the reference energy $E_0=130$ MeV. The black circles show the slice width $\sigma_M$ from the simulations.  The black dotted line at the left plot shows the results of the reconstruction with Eq.~(\ref{eq_main}). The other lines at this plot show contribution of different terms. For the data shown in the right plot the reconstruction is impossible. The lines show the expected values calculated using the true data of Table~\ref{T2}.} \label{Fig_fit}
\end{figure}

The results of the simulations are shown on Fig.~\ref{Fig_fit} and in the second row of Table~\ref{T2}. The left plot shows the results for the beam matched at each energy. The right plot shows the results for the beam matched only at the reference energy $E_0=130$ MeV. The black circles show the slice width $\sigma_M$ from the simulations.  The black dotted line at the left plot shows the results of the reconstruction with method of Section~\ref{sec2.1} using Eq.~(\ref{eq_main}). The other lines at this plot show contribution of different terms of Eq.~(\ref{eq_main}) as found from the reconstruction.

For the data shown in the right plot of Fig.~\ref{Fig_fit} the reconstruction was impossible. The black dotted line shows the expected values calculated by Eq.~(\ref{eq_main}) using the true data listed in the first row of Table~\ref{T2}. The other lines at this plot show contribution of different terms of Eq.~(\ref{eq_main}) as found from the true values of Table~\ref{T2}.

\begin{table}[htbp]
	\centering
	\caption{The true and the reconstructed data from the beam dynamics simulations at the reference energy $E_0=130$ MeV.}
	\label{T2}
	\begin{tabular}{lccccc}
		\hline\hline
			Parameter &\boldmath $\sigma_E$ &\boldmath $\sigma_I$&\boldmath$\sigma_B$&\boldmath$\sigma_R$&\boldmath$\epsilon_n$\\
			Units &keV&$\mu$m & $\mu$m& $\mu$m&$\mu$m\\
		\hline
		True values &$5.90$&$80.3$ &$35.4$&$28 $ &$0.53$\\
		Energy scan method &$5.89$& &&41 &\\
		Dispersion scan method &$5.97$&$81.8$ &$36.0$&$26.4 $ &$0.55$\\
		\hline\hline
	\end{tabular}
\end{table}

Thus, in the setup of the European XFEL the slice matching procedure should be applied on each step. However even with the matching the reconstruction could be non-accurate due to changes in the slice emittance and the energy spread at different energies. 

\subsection{Numerical modeling of the experiment based on dispersion scan }\label{sec3.3}

Another method for the slice energy spread measurement  proposed in the Section~\ref{sec2.2} uses the variation of the dispersion instead of the energy. To test the viability of the method, we carried out numerical experiments similar to that described above with two scans: dispersion scan and TDS voltage scan. 

The simulations have been done at the reference beam energy of 130 MeV for two different uncorrelated energy spreads: 5.9 keV and 2 keV. The first scan was performed with dispersion on the OTR screen of 0.6, 0.8, 1.0 and 1.2 m. The TDS voltage  was kept constant at 0.61 MV. The second scan was performed with TDS voltages of 0.47, 0.56, 0.65, 0.75, 0.84 MV. The dispersion  was kept constant at 1.2 m. 
 
\begin{figure}[htbp]
	\centering
	\includegraphics*[width=1\textwidth]{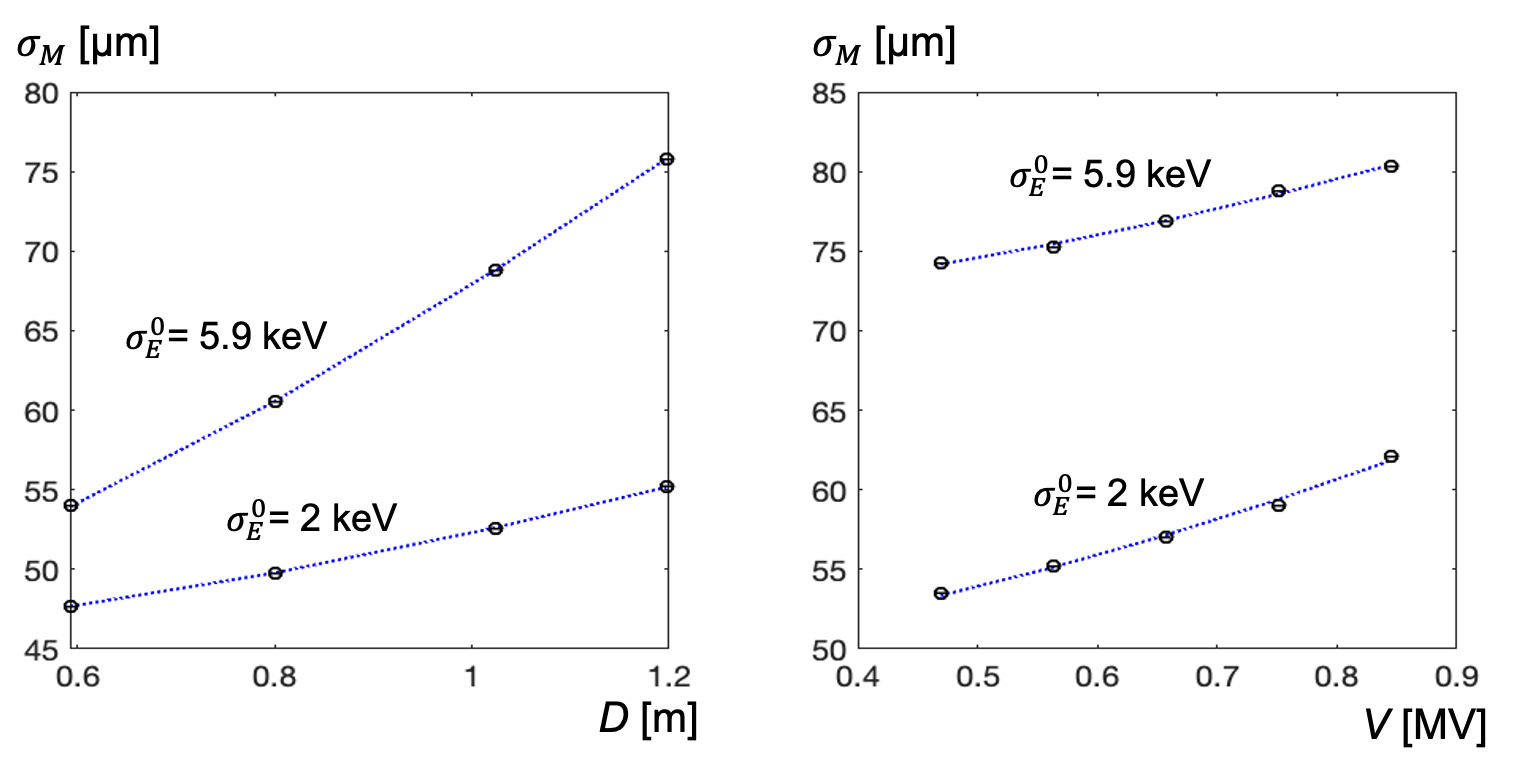}
	\caption{Beam size $\sigma_M$  (black circles) from  the dispersion scan and the TDS voltage scans as obtained from beam dynamics simulations. The blue dotted lines presents the curves of the numerical fit with Eq.(\ref{eq_sD}) and Eq.(\ref{eq_sV}).. } \label{Fig_size_disp}
\end{figure}

The results of the modelling are shown in Fig.~\ref{Fig_size_disp}. The black circles show the central slice width $\sigma_M$ obtained from beam dynamics simulations. The blue dotted lines presents the curves reconstructed by method of Section~\ref{sec2.2}.

Using reconstruction procedure described in Section~\ref{sec2.2}  for the true energy spread of 5.9 keV we got the reconstructed values listed in the last row of Table~\ref{T2}. We see that all values are reconstructed with high accuracy. For the true energy spread of 2 keV the reconstructed energy spread is 2.13 keV.  

Contrary to the first method based on energy scan in the second method the slice emittance does not change during the dispersion and the TDS voltage scans.

\section{Measurements}\label{sec4}

In the analysis of the images obtained in the experiment we have followed the same procedure as in the simulations. At each point we took 30 images and for each of them we calculated the mean slice energy and the slice size. The slice length was taken about 0.2 ps and the slice width was found by  fitting  to the Gaussian shape. Then the  slice width at the extremum of the mean slice energy curve has been taken as $\sigma_M$.  

From the measurements we estimate that the standard deviation error in $\sigma_M$ is below 1.5 \%.  Hence the error in the mean value from the 30 measurement is below 0.3\%.  

Due to substantial difficulties with the energy scan method in the experiment we change the order of consideration and consider the energy scan method at the beginning.

\subsection{Results obtained with the dispersion scan}\label{sec4.1}

The measurements presented in this section have been conducted at constant electron beam energy $E_0=130$ MeV. The parameters of the gun has been optimized to have small normalized slice emittance $\epsilon_n\approx 0.4$~$\mu$m.

\begin{figure}[htbp]
	\centering
	\includegraphics*[height=100mm]{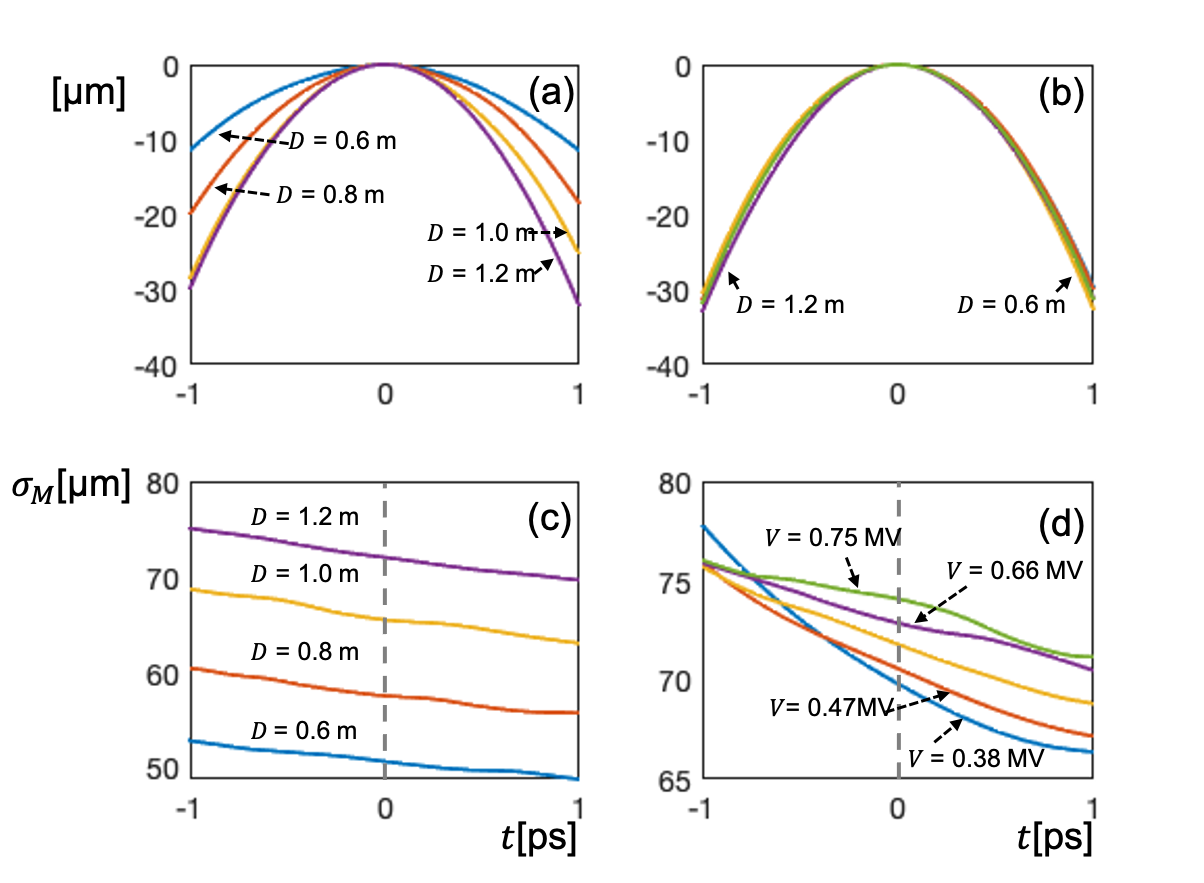}
	\caption{Measured curves with the dispersion scan method. (a) Mean vertical position of the slices on the screen along the bunch for different dispersion values. (b)  Mean vertical position of the slices on the screen along the bunch for different TDS voltages.  (c) Vertical size of the slices on the screen along the bunch for different dispersion values.  (b)  Vertical size of the slices on the screen along the bunch for different TDS voltages. The gray dotted lines present the position of the reference slice.} \label{Fig_ExMC}
\end{figure}

At the first step of the energy scan method we conducted the measurements at constant dispersion $D_0=1.181$ m with different  voltages $V$ of TDS. The voltages are listed in the first row of the Table~\ref{Table_VD}. The measured values of the slice width $\sigma_M$ together with the errors are listed in the second row of the Table.

At the second step we conducted the measurements at constant TDS voltage $V_0=0.61$ MV with different  dispersion values $D$ as listed in the third row of the Table~\ref{Table_VD}. The measured values of the slice width $\sigma_M$ together with the errors are listed in the last row of the Table.
 
\begin{table}[htbp]
	\centering
	\caption{Two first rows show the beam sizes measured at different TDS voltages. The last two rows present the beam sizes measured at different dispersion values.}
	\label{Table_VD}
	\begin{tabular}{lcccccc}
		\hline\hline
		\boldmath $V$ &MV & 0.375&	0.469 &0.563&0.657&0.751 \\
		\boldmath $\sigma_M$ &$\mu$m &$69.87\pm0.12$&	$70.64\pm0.10$&$71.86\pm0.13$&$72.85\pm0.17$&$74.12\pm0.14$ \\
		\hline\hline
		\boldmath $D$ &m & 0.578&	0.789 &1.006&1.181 &\\
		\boldmath $\sigma_M$ &$\mu$m &$50.62\pm0.08$&	$57.49\pm0.09$&$65.43\pm0.1$&$72.05\pm0.1$& \\
		\hline\hline
	\end{tabular}
\end{table}

Fig.~\ref{Fig_ExMC} shows the curves measured in the core of the electron bunch. At the position of the OTR screen the electron bunch has rms length of 4 ps.  Figures (a) and (b) show vertical position of the slice on the screen.  Figures (c) and (d) show the slice width on the screen. The gray dotted line defines the position of the reference slice. It is the extremum of the curves shown in plots (a) and (b). The same as in the simulations the reference slice do not have the minimal width but its position well defined by the extremum of the mean slice position curve.

\begin{figure}[htbp]
	\centering
	\includegraphics*[width=0.95\textwidth]{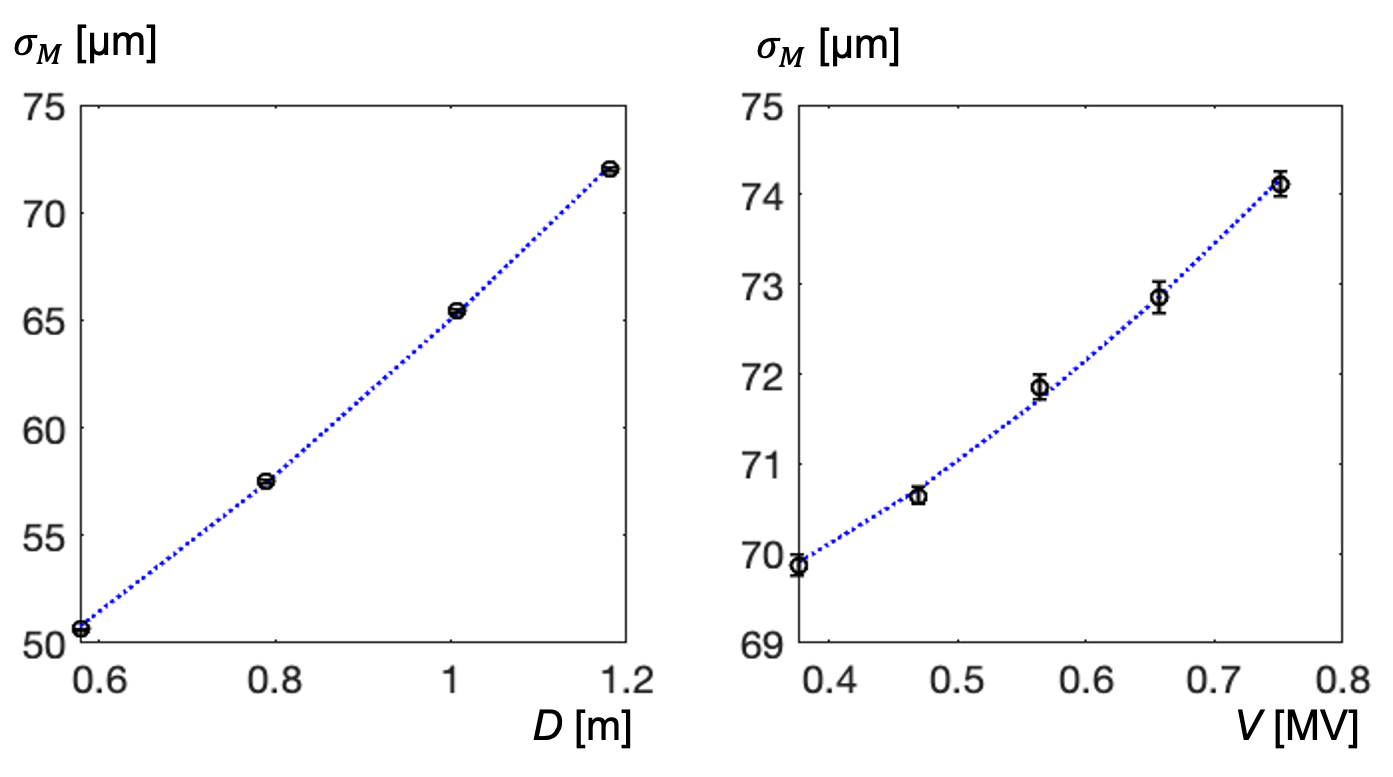}
	\caption{The black circles with error bars at the left plot show the measured slice width $\sigma_M$ for different dispersion values $D$.  The black circles with error bars at the right plot show the measured slice width $\sigma_M$ for different values of TDS voltage  $V$.  The blue dotted lines are obtained by the numerical fit to Eq.(\ref{eq_sD}) and Eq.(\ref{eq_sV}).} \label{Fig_ExDS}
\end{figure}

The measured values from Table~\ref{Table_VD} correspond to the coordinate $t=0$ ps at  Fig.~\ref{Fig_ExMC}. They are plotted in Fig.~\ref{Fig_ExDS} by black circles with error bars. The blue dotted lines are obtained by the numerical fit to Eq.(\ref{eq_sD}) (left plot) and Eq.(\ref{eq_sV}) (right plot).

The coefficients of Eq.(\ref{eq_sV}), Eq.(\ref{eq_sD}) obtained from the scans are listed in Table~\ref{Table_Coef}. In the same Table are presented the physical values of interest with the estimated errors. They are obtained with the help of Eqs.(\ref{eq_sall})-(\ref{eq_sB0}).  The errors are estimated by the numerical experiment described in Section~\ref{sec2.1}. 

If we take into account that the estimated instrumental errors in the setup of the TDS voltage and dispersion are smaller than 2 \% then we can state that the uncorrelated energy spread in the core of the beam is equal to $5.9\pm0.1$ k eV.

\begin{table}[htbp]
	\centering
	\caption{The reconstructed data from the measurements with the dispersion scan method.}
	\label{Table_Coef}
	\begin{tabular}{lcccccccc}
		\hline\hline
		\boldmath $A_V$ &\boldmath $B_V$ & \boldmath $A_D$ &	\boldmath $B_D$  &\boldmath $\sigma_E$ &\boldmath $\sigma_I$&\boldmath$\sigma_B$&\boldmath$\sigma_R$&\boldmath$\epsilon_n$\\
		$m^2$&$m^2/MV^2$ &$m^2$&	&keV&$\mu$m & $\mu$m& $\mu$m&$\mu$m\\
		\hline
		4.68e-9 &1.45e-9 & 1.75e-9&	2.48e-9 &$5.948\pm0.06$&$71.4\pm3$ &$31.4\pm1.3$&$27.6\pm1.5 $ &$0.42\pm0.02$\\
		\hline\hline
	\end{tabular}
\end{table}

 Note that the emittance estimation agrees well with the independent method of the mesurement of the beam emittance (see Fig.~\ref{Fig_ExEmit}). Finally the estimated screen resolution $\sigma_R$ agrees with the numbers published in ~\cite{OTR, Prat20}.
 
\subsection{Results obtained with the energy scan}\label{sec4.2}

\begin{figure}[htbp]
	\centering
	\includegraphics*[height=60mm]{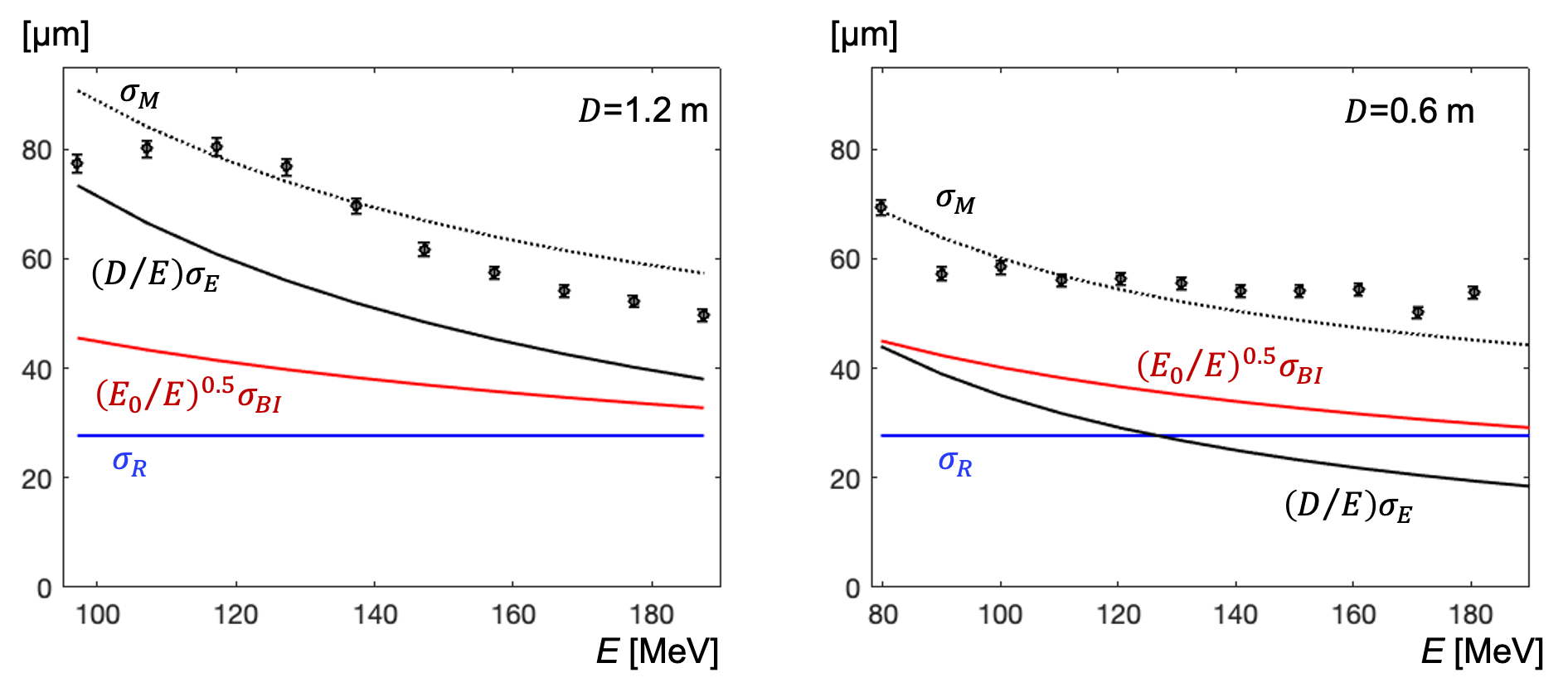}
	\caption{Comparison of the measurements of the energy scan (dots with error bars) with the values calculated from the data of the dispersion scan} \label{Fig_Ex10}
\end{figure}

We have done the energy scans with  dispersion values of 0.6 and 1.2 meters. The results are shown in Fig.~\ref{Fig_Ex10}. We were not able to do the reconstruction from the data measured and simply compare the measurements with the expected values calculated from the results of the previous method presented in Table~\ref{Table_Coef}. 

Taking into account the issues with the beam matching and non-constant slice emittance  (see Section~\ref{sec3.2}) we think that there is no contradiction between the data.

It is shown in Section~\ref{sec3.2} that the energy scan method could be used but requires stringent control of the shape of the longitudinal phase space and very accurate matching of the beam to the optics before TDS.  Unfortunately, we have not managed in two very time-consuming experiments to show it.

\subsection{Validation of the experimental results}\label{sec4.2}

In this section we consider several arguments to confirm the accuracy of the obtained data.

The energy spread estimation based on  Eq.(\ref{eq_sall}) uses only coefficient $A_V$ and $A_D$. But there is another equation
\begin{align}\label{eq_ESb}
	\sigma_E= \frac{E_0}{D_0}\sqrt{D_0^2 B_D- V_0^2 B_V}, 
\end{align}
based on two other coefficients, $B_D$ and $B_V$, from the numerical fits. From  Eq.(\ref{eq_ESb}) we obtain that the energy spread is equal to 5.946 keV that agrees with the previous estimation (see Table~\ref{Table_Coef}) with accuracy 0.03\%.

\begin{figure}[htbp]
	\centering
	\includegraphics*[height=65mm]{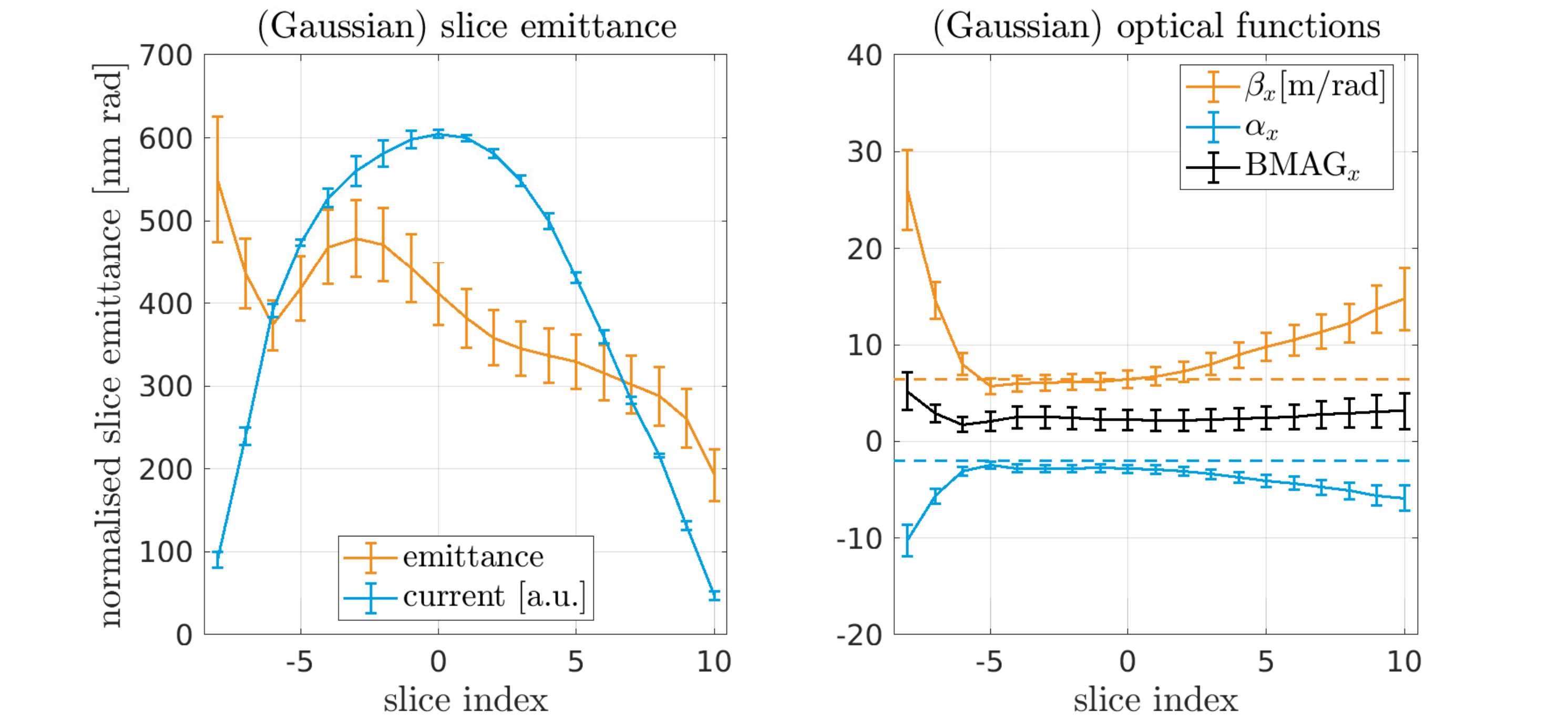}
	\caption{The slice emittance along the bunch measured by standard method~\cite{Bolko}. Black line on the right plot is BMAG parameter~\cite{Sands}. } \label{Fig_ExEmit}
\end{figure}

In order to check the estimation of the emittance  $\epsilon_n$ we have done an independent measurement of the slice emittance with the standard tools~\cite{Bolko} used by operators of the facility. The results of independent measurement of the slice emittance are shown in Fig.~\ref{Fig_ExEmit} and the emittance of the central slice (slice index 0) agrees with the value listed in Table~\ref{Table_Coef}.

We had additional possibility to do the measurement of the slice energy spread with the laser heater tuned for maximal SASE radiation energy. We have found that the energy spread in the electron bunch was $7.5\pm0.1$ keV. 

In theoretical studies of microbunching carried out by our colleague M. Dohlus (see, for example,~\cite{Dohlus}) the optimal energy spread  after laser heater for  microbunching suppression is nearly 8 keV. This number agrees reasonable with the measured one.

\section{Discussion}\label{sec5}

The theoretical calculations with different numerical models predict the uncorrelated energy spread below 1 keV. The discrepancy between the theoretical estimations and the measurements could be caused by neglecting of full physics in the simplified numerical models. For example, it could be that the emission process from the cathode should be simulated differently. Additionally we do not take into account the intrabeam scattering and wakefields in the RF gun cavity. The number of macroparticles used in the simulations does not allow to take into account the microbunching during the transport from the gun to the OTR screen. 

It was shown in~\cite{DiMitri, Huang} that the intrabeam scattering in the injector section increases the energy spread considerably and has to be taken into account. For example, a simple estimation of the induced energy spread due to IBS from ~\cite{Huang} reads
\begin{align}
	\sigma_{E}^{IBS}= \frac{2 r_e ^2 N_b}{\epsilon_n} \int{\frac{ds}{\sigma_x \sigma_z}},
\end{align}
where $r_e$ is the electron radius, $N_b$ is the number of electrons, $\sigma_z$ is the rms length of the bunch, $\sigma_x$ is the transverse rms size and integration is done along the bunch path $s$. If we use this equation with the parameters used in the paper we obtain that the energy spread introduced during the beam transport from the gun to OTR is about 2 keV. Hence it is considerable effect and should be taken into account in the simulations. We are considering now different models of IBS to include IBS in  the beam dynamics codes

The energy spread from the RF gun measured at the European XFEL for charge of 250 pC is $5.9\pm0.1$ keV. This number is approximately 3 times lower then the energy spread of $14.8\pm0.6$ keV reported recently by SwissFEL for the bunch charge of 200 pC~\cite{Prat20} .  The both guns use cesium telluride cathodes and the larger difference between these results requires additional efforts to understand. 

\section{Summary}\label{sec6}

We have described two methods for measurement of the slice energy spread of electron bunch. With the beam dynamics simulations we have identified substantial difficulties of the first method based on energy scan: we need match the beam and the slice emittance changes. The difficulties are confirmed in the real experiment. 

We have shown with the beam dynamics simulations and the measurements that the second method based on dispersion scan at the constant beam energy shows high accuracy and easy to conduct.

At the same time the measured slice energy spread of $5.9\pm0.1$ keV is several times higher than theoretically estimated and it requires additional theoretical research to clarify.

\section{Acknowledgement}\label{sec7}

The authors thank M. Dohlus  and M. Krasilnikov for helpful discussions. We thank members of the European XFEL team for providing help and conditions to carry out the measurements.

\end{document}